\documentclass{article}
\usepackage[english]{babel}
\usepackage[a4paper]{geometry}
\usepackage{amsmath}
\usepackage{physics}
\usepackage{graphicx}
\usepackage[colorlinks=true, allcolors=blue]{hyperref}
\usepackage{stackengine}
\usepackage[super,sort&compress,comma]{natbib}
\usepackage{caption}
\usepackage{subcaption}
\usepackage{mwe}
\usepackage{soul,xcolor}
\usepackage{authblk}
\usepackage{url}

\title{High-accuracy inference using HfO$_x$S$_y$/HfS$_2$ Memristors}

\author[1,2]{Aferdita Xhameni}
\author[1,2*]{Antonio Lombardo}

\affil[1]{London Centre for Nanotechnology, 19 Gordon St, London, WC1H 0AH, United Kingdom}
\affil[2]{Department of Electronic \& Electrical Engineering, Malet Place, University College London, WC1E 7JE, United Kingdom}
\affil[*]{Corresponding author: Antonio Lombardo (email: a.lombardo@ucl.ac.uk)}

\date{}

\begin{document}
\maketitle

\begin{abstract}
\noindent
We demonstrate high accuracy classification for handwritten digits from the MNIST dataset ($\sim$98.00$\%$) and RGB images from the CIFAR-10 dataset ($\sim$86.80$\%$) by using resistive memories based on a 2D van-der-Waals semiconductor: hafnium disulfide (HfS$_2$). These memories are fabricated via dry thermal oxidation, forming vertical crossbar HfO$_x$S$_y$/HfS$_2$ devices with an highly-ordered oxide-semiconductor structure. Our devices operate without electroforming or current compliance and exhibit multi-state, non-volatile resistive switching, allowing resistance to be precisely tuned using voltage pulse trains. Using low-energy potentiation and depression pulses (0.7V–0.995V, 160ns–350ns), we achieve 31 ($\sim$5 bits) stable conductance states with high linearity, symmetry, and low variation over 100 cycles. Key performance metrics—such as weight update, quantisation, and retention—are extracted from these experimental devices. These characteristics are then used to simulate neural networks with our resistive memories as weights. Neural networks are trained on state-of-the-art (SOTA) digital hardware (CUDA cores) and a baseline inference accuracy is extracted. IBM’s Analog Hardware Acceleration Kit (AIHWKIT) is used to modify and remap digital weights in the pretrained network, based on the characteristics of our devices. Simulations account for factors like conductance linearity, device variation, and converter resolution. In both image recognition tasks, we demonstrate excellent performance, similar to SOTA, with only $<$0.07$\%$ and $<$1.00$\%$ difference in inference accuracy for the MNIST and CIFAR-10 datasets respectively. The forming-free, compliance-free operation, fast switching, low energy consumption, and high accuracy classification demonstrate the strong potential of HfO$_x$S$_y$/HfS$_2$-based resistive memories for energy-efficient neural network acceleration and neuromorphic computing.
\end{abstract}

\textbf{Keywords}: \textit{potentiation, depression, memristors, resistive random access memory (RRAM) neuromorphic computing, hafnium disulfide, hafnium oxide, 2D layered material (2DLM), forming-free, compliance-free, van der Waals semiconductors}\leavevmode\newline

\section*{Introduction}
The economic and environmental cost of training and deploying neural networks for machine learning and artificial intelligence must be addressed \cite{mehonic2022masterplan} \cite{jones2018stop}. From their conception, neural networks have taken inspiration from the brain to enable and improve performance in machine learning tasks. Arrangements of artificial neurons and synapses comprise the network, where the strengths of connections between different nodes in the network (artificial neurons) are represented by the weight values of branches connecting the layers (artificial synapses). Once a network has been trained to solve a particular task, its weights encode the network's ability to evaluate new, untested data, and thus should be trainable, precise, and resilient to repeated programming and device ageing. In the earlier days of machine learning, computation for neural networks was performed on digital hardware such as central processing units (CPUs) that resulted in performance increases over time which generally followed Moore's law. However, since 2012, computation for machine learning on digital hardware has been performed on graphical processing units (GPUs) from which a doubling of performance has been achieved every 3.4 months or fewer. Aside from improving algorithms and the increased parallelism offered by GPU cores, this rapid increase in performance can also be explained by the rate at which GPU hardware has improved, with NVIDIA GPUs improving in computational performance by a factor of 317 since 2012 \cite{mehonic2022masterplan}. However, despite recent advances in more efficient algorithms and hardware architectures, a rethinking of machine learning systems at the most fundamental level is urgently required to address the ever-growing demand for computing power \cite{mehonic2022masterplan}. Taking inspiration from the brain, developing hardware that can co-locate processing and memory functions is key to breaking the von Neumann bottleneck which limits computational efficiency by necessitating the shuttling of data back and forth between processing and memory units \cite{Phys4NeuroComp}.\\

Many different types of devices pose as potential candidates for accelerating performance in machine learning tasks and surpassing the von Neumann bottleneck. Memristors are one class of simple, two-terminal analog devices which have shown promise in hardware acceleration for neuromorphic computing and machine learning tasks when integrated in densely-packed crossbar arrays \cite{adnanmemristorsreview}. In analogy with biological systems, where the transmission strength of incoming signals can be controlled at a synapse, most memristors can variably impede the flow of current due to the modulation of their conductance between multiple states. Conductance states in a memristor can be modified by application of electrical stress, such as a voltage or current pulse, where increasing the memristor's conductance state is referred to as potentiation, and decreasing its conductance state is  referred to as depression. Similarly to biological synapses, some memristors can retain programmed conductance states when electrical stress is removed, making them non-volatile memory devices. Hence, in most implementations where memristors are used for machine learning, the devices are integrated in crossbar arrays utilised for weight storage and update. The analog crossbar array can then be interfaced with digital integrated circuits via analog-to-digital converters for other processes in a machine learning task such as applying activation functions. In existing digital hardware, weight values are calculated and stored in separate logic and memory units, respectively. However, analog crossbar arrays of memristors offer vastly increased parallelism and avoid shuffling data back and forth as weight values  can be both programmed and stored as non-volatile conductance states in the same memristive hardware. Therefore, by inputting voltages across the rows of a crossbar array where memristors have been programmed to precise conductance states, and measuring the output currents along the columns, a memristor-based crossbar array can perform multiplication and accumulation operations (using Ohm's and Kirchhoff's laws) to enable fast matrix-vector multiplications (MVM). In some applications such as compressed sensing, this presents a key advantage of memristor-based crossbar arrays, which is to allow for different matrix-vector-multiplication (MVM) operations to execute in the same amount of time, regardless of the input data size (O(1) time complexity) \cite{adnanmemristorsreview}. The same is not true for GPUs, in which execution time grows with n$^2$ for input data of size n in MVM operations. A good example of the performance enhancements offered by memristive hardware in machine learning tasks can be found in the work of Yao et al. \cite{FullyHardwareMemrisML}. In their work, network weights were first trained on digital hardware, then transferred to physically-implemented crossbar arrays of TiN/TaO$_x$/HfO$_x$/TiN memristors, with modifications made to network weights to be aware of the characteristics of their memristor hardware \cite{FullyHardwareMemrisML}. Their memristor-based neural network was able to correctly classify a large proportion of previously unseen handwritten numbers from the MNIST dataset \cite{MNISTDataset}, resulting in a classification accuracy of 96.19$\%$, close to a digital hardware baseline score of 97.99$\%$ \cite{FullyHardwareMemrisML}. The small difference in accuracy, but significantly decreased energy consumption and improved performance density compared to using conventional digital hardware for storing weights clearly demonstrates a use-case for integrating memristor-based hardware accelerators with digital components. Therefore, owing to their potential for efficiency, scalability and strong non-volatile memory performance, memristors are a strong candidate for use as analog weight storage in crossbar arrays \cite{FullyHardwareMemrisML} \cite{zhu2023hybridhBNCMOS}.\\

Despite their excellent performance, there are a number of challenges associated with using memristive hardware for machine learning applications. Noise during conductance update or read steps can originate from a variety of sources, and reduces the effectiveness of using memristors for weight storage or weight update in a neural network. Precisely programming and distinguishing states in noisy devices which show highly non-linear conductance update within a limited conductance range can become impossible, leading to reduced machine learning accuracy. On the other hand, linear conductance update can allow for different conductance states to be more accurately distinguished during potentiation and depression, provided that there is low cycle-to-cycle variation and little drift in the devices. When trying to program memristors to represent specific weights in a neural network, this behaviour would facilitate a high machine learning accuracy. Furthermore, not only should conductance update be linear in potentiation and depression, but the device should also show a high degree of symmetry in both schemes. To enable the use of energy efficient and scalable memristive hardware in machine learning tasks, tailored potentiation and depression pulsing schemes in which pulse widths and amplitudes can be modulated by pulse number are often introduced  \cite{NonIdealON/OFFMemris} \cite{DealingMemrisNonIdeal} \cite{lu2021exploring} \cite{nguyen2021memristor}. Despite increasing computational latency and circuit area, these schemes not only linearize the conductance update with respect to programming pulse number, but can also reduce cycle-to-cycle programming noise too \cite{DealingMemrisNonIdeal}.\\

Another set of challenges associated with implementing memristive hardware in neural networks are the requirements of electroforming and current-compliance for each device. Electroforming is a one-time initialisation step, typically required by a class of memristors called resistive random access memory (RRAM) devices, based on insulating metal-oxides. RRAM devices have shown otherwise excellent performance, reliability, energy efficiency and scalability. However, requiring electroforming provides a barrier for their adoption since it necessitates increased peripheral circuitry and harms scaling \cite{ZnOformingfree}. Current compliance circuitry is required by most memristive devices which do not have a self-limiting mechanism to prevent high currents from damaging the device. In most implementations, memristive chips require 1-transistor-1-memristor architectures (1T1M) which limit integration density and increase computational complexity \cite{SnOFormingCompFree} \cite{ReRAMFormingFreeThin}. Defect engineering can be used to address electroforming by introducing a large number of defects or modifying the microstructure of the pristine device by providing pre-existing conductive pathways. Self-limiting currents in the memristor stack has been achieved by intentionally engineering heterostructures which slow the motion of charge carriers responsible for changing the device resistance such as oxygen vacancies \cite{SnOFormingCompFree} \cite{ReRAMFormingFreeThin}. One class of materials which allows for precise defect engineering and facile control of heterostructures with pristine interfaces is that of two-dimensional layered materials (2DLMs). Memristors based on partially oxidised 2DLM semiconductors have shown promise in energy efficiency, fast, compliance-free and electroforming-free operation \cite{AzizGaSMemristors} \cite{xhameni2025forming} along with strong non-volatile memory characteristics and independence from surrounding conditions such as water vapour, temperature and oxygen \cite{xhameni2025forming}.\\ 

While there are a variety of candidates and methods for integrating memristors in neural networks which address many of these issues, it is necessary to have a means of rapidly and accurately evaluating the potential performance of new classes of memristive devices for hardware accelerators. Such devices may be able to provide a breakthrough in performance which could carve a niche for their use within machine learning and neuromorphic applications. However, due to the nature of device fabrication with novel materials and engineering methods, the road from individual device fabrication and performance evaluation to wide-scale chip fabrication and programming is often long, expensive and does not allow for exploring the materials and parameter space for device optimisation, limiting adoption by industry. The Analog Hardware Accceleration Kit (AIHWKIT) developed by IBM \cite{IBMaihwkitRef} is a Python-based, open-source library which enables estimation of the performance of analog memristive hardware in a variety of machine learning tasks by implementing a wide range of measured device performance parameters, non-idealities and necessary peripheral circuitry in machine learning simulations \cite{UsingIBMaihwkit}. While performance estimations based on AIHWKIT may not take into account all possible challenges that may arise when exploring new hardware for analog in-memory computation, it allows for devices based on less-mature technologies to demonstrate their potential performance in memristive chips. Consequently, it provides an excellent route towards rapid device optimisation and materials screening without the need for the complex fabrication of large arrays of devices.\\

In this work, we investigate the machine learning performance of a simulated network/crossbar chip whose elements are experimental HfO$_x$S$_y$/HfS$_2$ memristors. Such devices have shown high potential for ML applications as they combine sub-nJ switching, excellent thermal and environmental stability, current self limiting (compliance free) and forming-free operation \cite{xhameni2025forming}. These devices were investigated with tailored potentiation and depression pulses, producing highly linear and symmetric conductance update with low cycle-to-cycle variation. Such linearity, together with the non-volatility of the states, enables the use of our memristors in neural networks to store synaptic weights. The network weights are mapped into a number of programming pulses and stored in the memristors as resistance or conductance values. As a result, despite accounting for a range of measured device characteristics, such as ON/OFF ratio, cycle-to-cycle and device-to-device variation, our simulations show high accuracy classification scores with the MNIST dataset \cite{MNISTDataset} and the more challenging CIFAR-10 dataset \cite{CIFAR10Dataset}, nearing SOTA performance. Combined with the high performance of the individual memristors we investigated, their potential performance in memristive chips for machine learning tasks provides strong motivation for further research in physical memristive chips based on this class of device.\\     

\section{Evaluation of ML performance of novel analog memories}
\begin{figure}[h!]
\begin{center}
\includegraphics[scale=0.3,trim=10 6 6 6,clip]{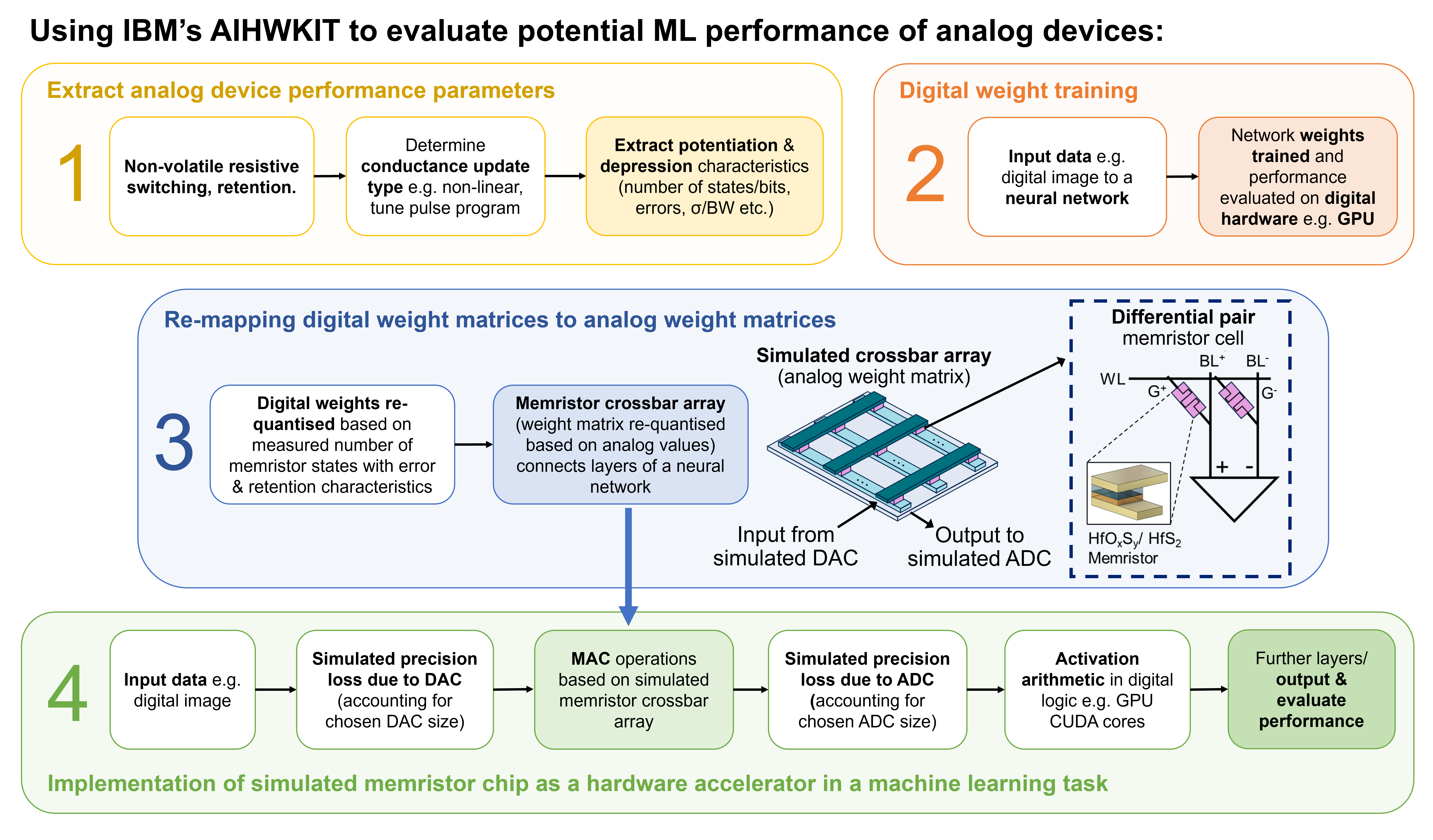}
\end{center}
\caption{Process flow for using the AIHWKIT \cite{IBMaihwkitRef} to evaluate potential machine learning performance of an analog memory device intended for use in a crossbar array for weight storage or update.} 
\label{fig1}
\end{figure}

To evaluate the potential machine learning performance of our devices in a suitable task, we have used an open source Pytorch toolkit developed by IBM called the Analog Hardware Acceleration Kit (AIHWKIT) \cite{IBMaihwkitRef}. In Fig. \ref{fig1}, we outline the typical workflow for evaluating the use of memristors in a simulated crossbar array used for a machine learning task \cite{UsingIBMaihwkit}.\\

Prior to using the toolkit, the performance of the device as a non-volatile memory should be evaluated, which involves testing the resistive switching of the device with voltage pulses and testing the non-volatility or retention characteristics of the programmed states. When designing a crossbar array of memristors for weight storage and update in a machine learning task, the conductance or resistance state of the device should then be programmed to increase and decrease in a linear and gradual manner (potentiation and depression, respectively.) Since weights held in digital logic can take values between -1 to 1, and conductance states represent weights in memristive hardware accelerators, often a differential configuration of memristors (Fig. \ref{fig1}, panel 3) is employed as only positive conductances can be encoded in each individual device. Therefore, each memristor cell is composed of two memristors in the array, with one corresponding to positive weights and the other corresponding to negative weights. An example of the measurement and data extraction process is shown in the next section.\\

Once device characteristics have been determined, a machine learning task should be chosen, and an associated neural network can be programmed and evaluated in its ability to solve the task. By default, both training and inference will run on digital hardware such as CPU or GPU cores, locally or by utilising cloud computing services. In our case, we ran our code locally on an NVIDIA RTX 3080 GPU, utilising its compute unified device architecture (CUDA) cores.\\

Using the device conductance update and retention characteristics, a suite of device features and non-idealities can be configured for the digital hardware to simulate while solving the machine learning task. These include but are not limited to: the device type (for example PCM, RRAM, etc), the number of conductance states/ bit resolution, voltage drops across rows and columns of the crossbar array due to interconnect resistance (IR drop), retention characteristics, update linearity and asymmetry and peripheral circuitry features such as ADC/ DAC size. In the context of simulating a crossbar array of memristors to store and update weights, all of these limitations and features in device performance are applied by modifying the way that weights (programmed in digital hardware) would change if the network were deployed onto analog hardware. Then, when the network deployed on simulated analog hardware is used for inference, the impact of programmable bit resolution, inaccurate weight programming, and limited retention on machine learning performance due to analog hardware characteristics can be evaluated by observing a difference in performance compared to the unmodified, SOTA digital hardware. It is worth mentioning that the network can also be trained on the simulated analog hardware \cite{KimEtal}. In this case, step 3 in Fig. \ref{fig1} is performed before step 2, such that the training is performed utilising the performance parameters and non-idealities of the analog hardware. In the following sections, we explore several examples of evaluating the machine learning potential for memristive hardware based on our experimental HfO$_x$S$_y$/HfS$_2$ memrsitors.\\

\section{Potentiation and Depression}
\begin{figure}[h!]
\begin{center}
\includegraphics[scale=0.52,trim=6 6 6 6,clip]{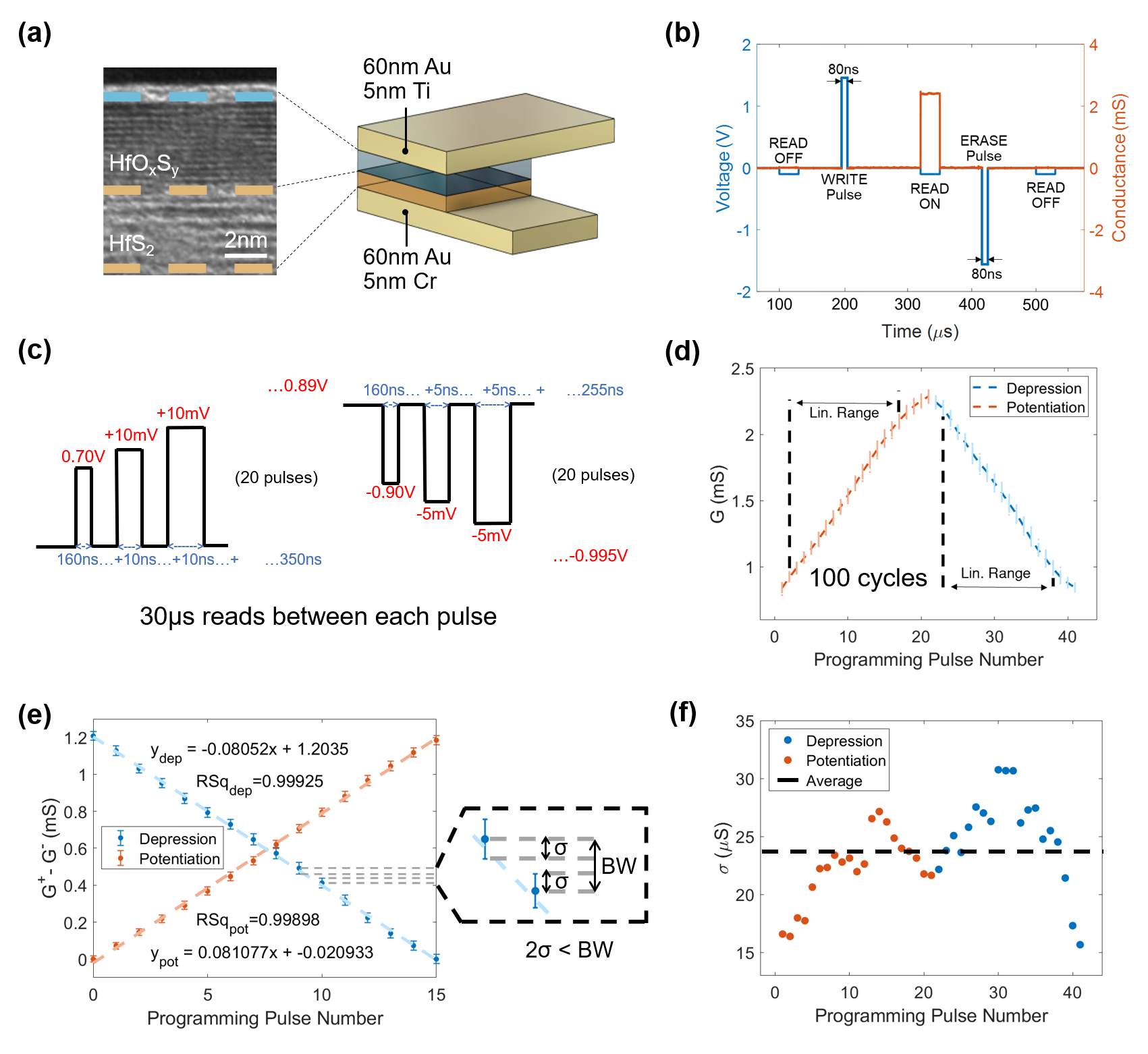}
\end{center}
\caption{(a) Forming-free, compliance-free device used in potentiation and depression experiments shows (b) robust non-volatile resistive switching with fast (80ns) WRITE and ERASE pulses. (c) We employ a tailored pulsing scheme to achieve gradual and linear conductance update which is required to attain the desired bit-resolution for high accuracy machine learning with our devices. (d) Distribution of conductances achieved using the pulsing scheme in (c), over 100 cycles. (e) Linear region extracted from (d), with average conductances and standard deviation fitted with linear functions. A high degree of symmetry between potentiation and depression and low standard deviation at all conductance values (f) allows for 31 bins (unique values of G$^+$ - G$^-$) to be defined with bin width (BW) $>$ 2$\sigma$. } 
\label{fig2}
\end{figure}

We extract relevant parameters for simulating image recognition performance of compliance-free and forming-free memristors based on a crystalline 2DLM semiconductor (HfS$_2$) which was partially dry oxidised to form the HfO$_x$S$_y$/HfS$_2$ structure shown in Fig. \ref{fig2}a. Devices were measured on a FormFactor MPS150 probe station, connected to a Keysight B1500A Parameter Analyzer with remote sensing units and B1530A WGFMU (waveform generator/ fast measurement unit) with a temporal resolution of 10ns. The devices show stable non-volatile resistive switching when measured with fast voltage pulses (Fig. \ref{fig2}b).\\

Similar to other RRAM technologies \cite{NonIdealON/OFFMemris} \cite{DealingMemrisNonIdeal} \cite{IBMaihwkitRef} \cite{lu2021exploring} \cite{FullyHardwareMemrisML} \cite{panetal}, our devices show non-linear conductance update characteristics when biased with repeated identical voltage pulses. However, although not ideal, this has been circumvented by using pulses with increasing voltage and pulse width (Fig. \ref{fig2}c) at a cost to increasing the required peripheral circuitry in a physical implementation of such a circuit \cite{NonIdealON/OFFMemris} \cite{DealingMemrisNonIdeal}. 20 pulses were used for both potentiation and depression to leave headroom to extract an optimal performance range, with one complete potentiating pulse train and one complete depressing pulse train constituting one programming cycle, therefore containing 40 programming pulses. To ensure robust characterisation of our devices, read pulses were employed as -0.1V, 30$\mu$s pulses, 20$\mu$s apart from programming pulses, avoiding any contribution to read currents from spurious charging or discharging capacitances due to the high frequency operation. The voltages and pulse widths employed were low ($<$1V and $<$350ns respectively) and are indicated in Fig. \ref{fig2}c.\\

Potentiation and depression pulse trains were conducted on a single device for 100 cycles to determine the resilience of the device to repeated programming (Fig. \ref{fig2}d). The raw conductance read data for each of the 100 cycles is shown superimposed on one complete programming cycle, with the average values plotted using a dashed line. From the data, we extract the conductance states obtained from potentiation and depression within the most linear range of both regimes (Fig. \ref{fig2}d). We require a differential configuration of our devices to represent positive and negative weights, therefore weight values (w) encoded by our devices must be represented by the difference in conductance of the memristors on a positive (G$^+$) and negative (G$^-$) branch (w $\propto$ G$^+$ - G$^-$). In this configuration, each unique combination (G$^+$ - G$^-$) of quantised conductance states is assigned a "bin" number, and the separation between neighbouring bins is defined as the bin-width. This results in 31 total bins (or log$_2$(31)$\sim$5 bits) being accessible for reliable programming in both potentiation and depression combined (Fig. \ref{fig2}e). We choose to program our devices to $\sim$5 bits as this is the point at which the precision of analog implementations can be superior to digital ones while not having much higher multiply-and-accumulate (MAC) energy and is therefore a realistic application for our devices \cite{5bitsForAnalogAI}.\\ 

Additionally, the cycle-to-cycle variation and consequently the standard deviation of each state is also crucial for determining how reliably a memristor within a memristive chip can achieve a predicted or specified conductance when programmed with an associated pulse train. Within our linear range, each step between bins or average conductance states is encoded by the average bin-width 81.077$\mu$S or 80.520$\mu$S for programming in potentiation or depression, respectively. Crucially for machine learning accuracy, the standard deviations of neighbouring states do not overlap. This is indicated between a pair of neighbouring states in Fig. $\ref{fig2}$e. Furthermore, comparing the gradients of the fitted lines for both potentiation and depression, we observe only a very small difference. A high degree of symmetry between potentiation and depression conductance update also positively influences machine learning accuracy and is present in our data. The high R-squared values for both lines (0.99898 for potentiation and 0.99925 for depression) also indicate how closely we can fit a linear conductance update model to our data, from which we will base our simulated crossbar array devices for machine learning. Fig. $\ref{fig2}$f shows the standard deviation of each state, which can be taken at an average value of 23.700$\mu$S. Overall, the device shows strong linear and symmetrical conductance update characteristics at low energy (23.74nJ $\pm$ 1.26nJ total programming energy per complete potentiation/depression cycle, averaged over 10 cycles), without requiring electroforming or current compliance, from which we can build a device model for simulating machine learning performance of a memristor chip.\\

\section{MNIST image classification}
\begin{figure}[h!] 
\begin{center}    
\includegraphics[scale=0.53,trim=4 4 4 4,clip]{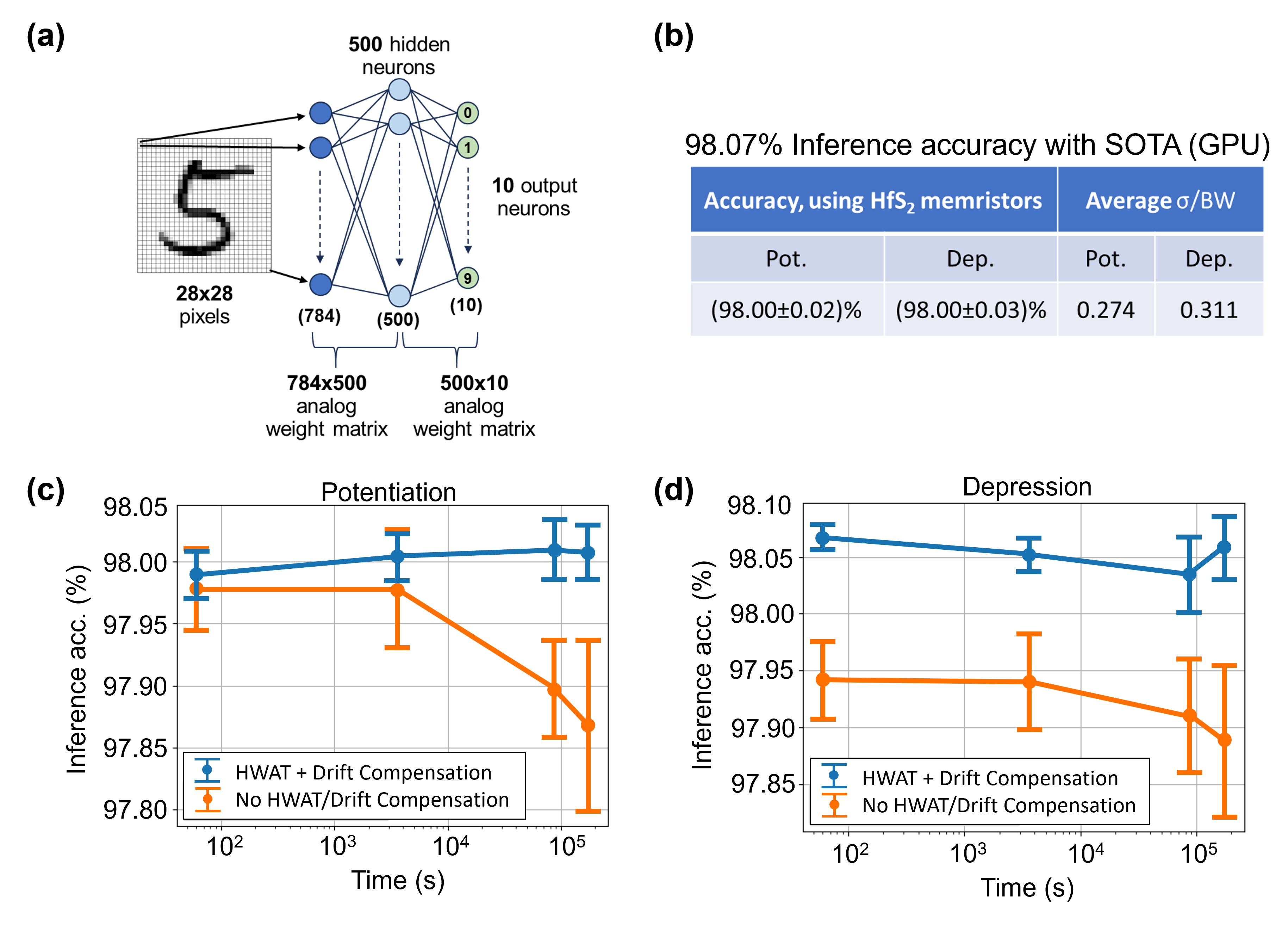}  
\end{center}
\caption{ (a) Network architecture employed for MNIST image classification task. (b) High accuracy inference results with 5 repetitions indicating consistent MNIST classification performance close to SOTA devices. Resilience of memristive hardware to drift in potentiation (c) and depression (d) given global drift compensation and hardware-aware re-training of the network weights.}
\label{fig3}
\end{figure}

In Fig. $\ref{fig3}$a, we show the network utilised to classify handwritten number images included in the MNIST dataset \cite{MNISTDataset}. The 28x28 input images are flattened to a 1x784 vector in the first layer of the network. We use one hidden layer consisting of 500 neurons, and the network has a fully connected architecture where each neuron from one layer is connected to all neurons in the subsequent layer. We use this architecture as it has been shown to result in high accuracy classification of handwritten numbers in the MNIST dataset \cite{MlpMnistStructure}. Finally in the last layer of the network in Fig. \ref{fig3}a, a predicted handwritten number is determined from 10 possible values (0 to 9). Deploying this network architecture on SOTA digital hardware (an NVIDIA RTX 3080 GPU) and training the network weights for 30 epochs results in an inference accuracy of 98.07$\%$.\\

In the simulated analog implementation of this network, we employ a differential configuration of two memristors (one to represent positive weights and the other for negative weights) between each neuron, as described in the previous secion, and in Fig. \ref{fig1}, panel 3. When simulating deployment of the network on our memristive hardware, we map network weights to the number of programming pulses required to reach each corresponding analog weight within the linear operating range of the device, using the linear fit in Fig. \ref{fig2}e. The ‘0’ weight value is mapped to the start of the device linear range.\\

Based on the electrical data in Fig. $\ref{fig2}$, we extracted relevant performance parameters pertaining to (i) cycle-to-cycle variation, (ii) linearity, (iii) symmetry, (iv) IR drop, (v) bit-resolution and (vi) the characteristics of conductance update in our HfO$_x$S$_y$/HfS$_2$ devices which inform our device model. However, prior to deploying weights to our simulated arrays of HfO$_x$S$_y$/HfS$_2$ devices, we perform a further 5 epochs of hardware-aware training (HWAT) on the network. During HWAT, the network learns to ensure robust weight deployment to our HfO$_x$S$_y$/HfS$_2$ devices by retraining network weights on digital hardware for a small number of cycles while accounting for the characteristics of the analog hardware which we wish to deploy the network on. Although we have a good picture of device characteristics over a range of experimentally measured parameters (i-vi above), in lieu of data with similar statistical significance, we have simulated the impact of device-to-device variation informed by literature on arrays of HfO$_2$ memristors fabricated by a scalable method (ALD) \cite{HfO2DTODVar} and arrays of hBN-based 2D layered memristors \cite{hBNUltraFastRESETCurveBump}. This has informed a baseline value of 30$\%$ for conductance update and how reliably we can achieve the minimum/ maximum conductance states of our devices. Therefore, during HWAT, we account for our experimentally measured analog hardware characteristics (i-vi listed above) and for simulated device-to-device (DTOD) variation. Inference accuracy is subsequently extracted by simulating deployment of the HWAT-modified weights on our HfO$_x$S$_y$/HfS$_2$ devices, and evaluating the proportion of correctly predicted handwritten numbers from an unseen test set from the MNIST dataset.\\ 

Simulations were conducted for both potentiation and depression, with separate noise characteristics corresponding to each programming mode. In both potentiation and depression (positive and negative weight update, respectively) we achieve 98.00$\%$ accuracy with low variation across 5 runs, only 0.07$\%$ lower than SOTA accuracy, showing the potential of this hardware to solve machine learning tasks with high accuracy (Fig. \ref{fig3}b). This is largely attributed to the linear operation and low $\sigma$/Bin-Width (BW) ratio in both programming regimes extracted from Fig. \ref{fig2}e.\\

In our simulations, we have also utilised the capability of the AIHWKIT to simulate the peripheral circuitry connecting analog memristor hardware and SOTA hardware, as memristive chips cannot be operated in isolation. We map analog weights to digital values and utilise 8 bit ADC and DACs which, despite being higher resolution than our devices which have shown $\sim$5 bit operation, make our circuit more resilient to programming noise albeit at a cost to total computation energy and chip area. Along with adaptive scaling of input data for the first few batches of training data (ensuring weights are represented as accurately as possible within the limitations of our analog hardware), this ensures that our simulation is not agnostic of the other peripheral hardware required to operate a memristive chip for use in a neural network for machine learning tasks.\\

To create an even more complete device model, we also utilised conductance drift data from the ReRAMWan2022 analog device model \cite{ReRAMWan2022}. This model is based on electrical measurements of thousands of HfO$_2$ memristors, allowing us to reasonably predict inference accuracy drift for a future array of our own HfO$_x$S$_y$/HfS$_2$ devices fabricated by a scalable method. Over the same time span for which we observe $<$3$\%$ conductance drift across the whole range of conductances programmed in our devices \cite{xhameni2025forming}, the ReRAMWan2022 data shows considerably more conductance drift, up to 30$\%$ \cite{ReRAMWan2022}, which we take as worst-case scenario baseline due to its higher statistical significance. We compare the retention of inference accuracy by simulating deployment on our HfO$_x$S$_y$/HfS$_2$ devices with HWAT and drift compensation against a baseline model, which we take to mean without HWAT and without any compensation for conductance drift described by the ReRAMWan2022 model \cite{ReRAMWan2022} (Fig. \ref{fig3}c-d). This shows that the impact of HWAT on MNIST inference accuracy is relatively small for our devices, and therefore may be unnecessary. However, by applying global drift compensation, we observe superior retention of inference accuracy over the tested period compared to the baseline ReRAMWan2022 drift model \cite{ReRAMWan2022} without compensation.\\ 

\section{CIFAR-10 image classification}
\begin{figure}[h!] 
\begin{center}    
\includegraphics[scale=0.255,trim=250 705 250 190,clip]{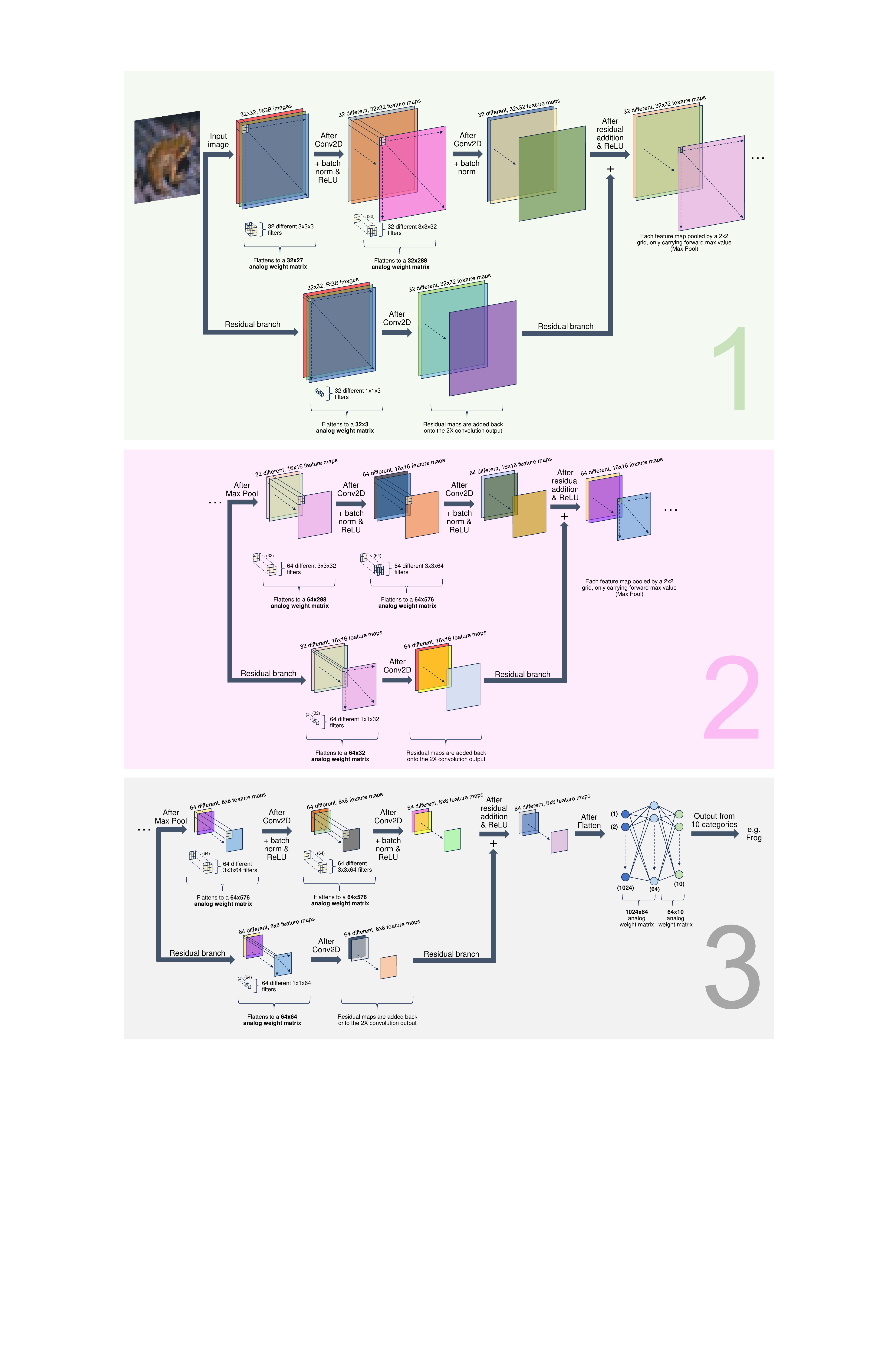}
\end{center}
\caption{Convolutional neural network composed of 3 ResNet blocks and a fully connected layer. This network was trained with data from the CIFAR-10 dataset augmented with random horizontal flips, rotations, normalisation, resizes and crops. Simulated arrays of HfO$_x$S$_y$/ HfS$_2$ device arrays act as analog weight matrices in the network.} 
\label{fig4}
\end{figure}

To further evaluate the potential of our devices for machine learning applications, we chose an image classification task based on the CIFAR-10 dataset \cite{CIFAR10Dataset}. This dataset contains 60,000 32x32 pixel RGB images of 10 different categorical items, including dogs, cats, frogs and others. To classify the CIFAR-10 dataset, we implemented a convolutional neural network (CNN), shown in Fig. $\ref{fig4}$. This network is composed of 3 repeated blocks, where the input image is split into two branches. In the upper branch, it undergoes two convolutions-being scanned by a 3x3 filter to extract a number of reduced-dimension feature maps each time. After the first convolution in the upper branch, the feature maps are normalised with respect to a dynamically calculated mean and standard deviation (batch normalisation, improving stability during training) and then passed through a rectified linear unit (ReLU) activation function which introduces non-linearity in the data, improving the ability of the network to learn complex patterns. After the second convolution in the top branch, only a further ReLU operation is performed.\\

In the lower branch, only one convolution is performed, with a 1x1 filter so as to extract an equivalent number of feature maps to the other branch. This lower branch corresponds to the residuals which are then combined with the result of the two convolutions in the upper branch, and pooled to the maximum value in a 2x2 filter which is passed over the different channels, to reduce the dimensionality of the output of each block. The output of each preceding block becomes the input of the next. These blocks are one possible implementation of ResNet blocks which have been shown to be successful in image recognition machine learning tasks \cite{ResNET1stpaper}. We employ only 3 ResNet blocks, although many modern network architectures implement 10s of these blocks to achieve very high accuracy in even more challenging tasks. However, this comes at a cost to the number of trainable weights increasing, thus we employ only 3 blocks to maintain the number of memristors required to implement this network relatively low. The final part of the network is a fully connected network with 3 layers, outputting a prediction from the 10 possible categories. Given a differential configuration of memristors, in total, $\sim$420,000 HfO$_x$S$_y$/ HfS$_2$ devices would be required to store $\sim$210,000 trained weights for the whole network, due to convolutions and other operations in each of the three blocks (see analog weight matrix dimensions in each panel, Fig. \ref{fig4})\\

\begin{figure}[h!]
\begin{center}
\includegraphics[scale=0.48,trim=6 6 6 6,clip]{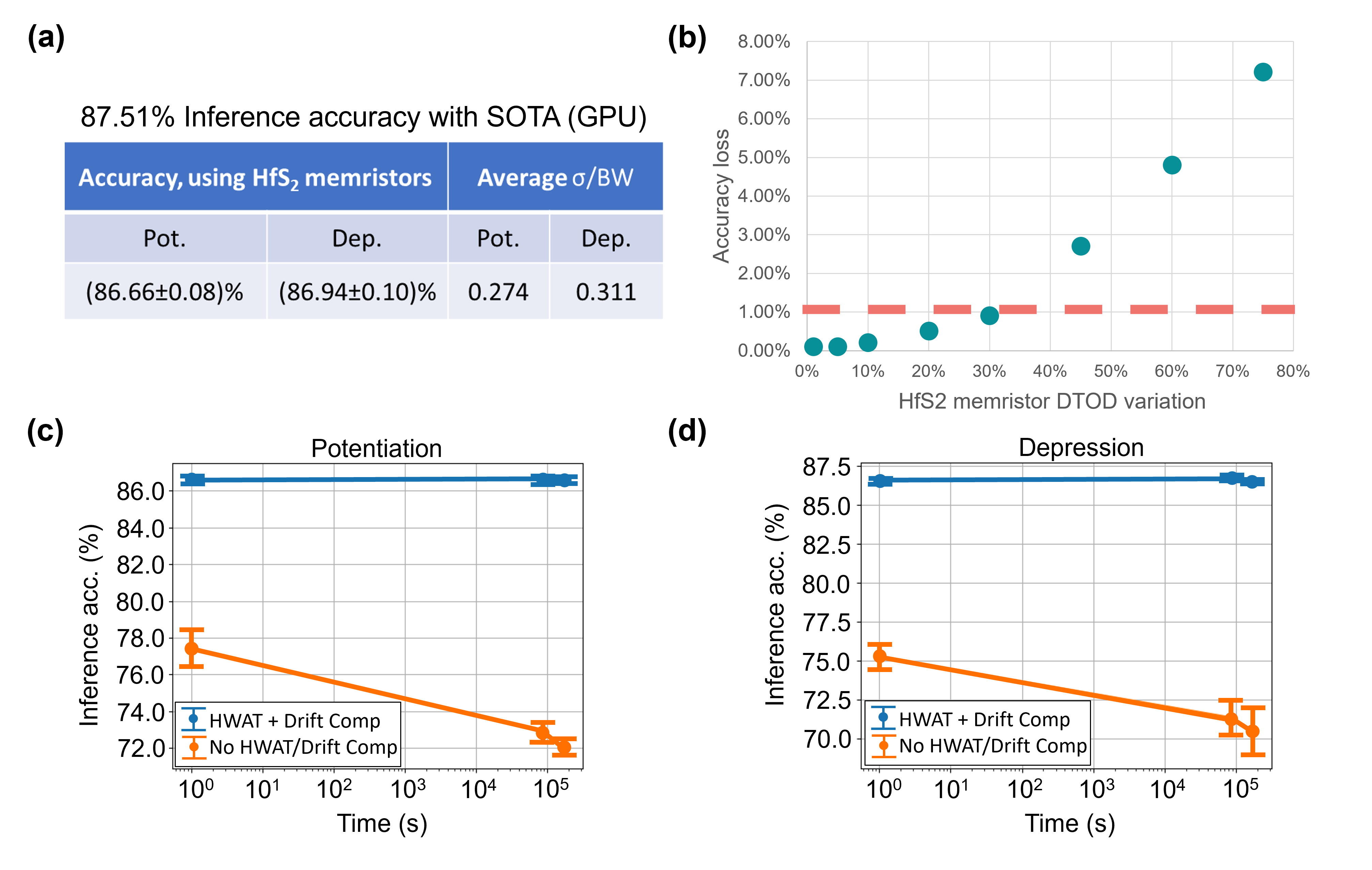}
\end{center}
\caption{(a) High accuracy, low variation inference results with 5 repetitions indicating consistent CIFAR-10 classification performance close to SOTA devices. (b) Device-to-device (DTOD) variation below 30$\%$ results in $<$1$\%$ drop in inference accuracy compared to SOTA devices. Resilience of memristive hardware to drift in potentiation (c) and depression (d) given global drift compensation and hardware-aware re-training of the network weights.} 
\label{fig5}
\end{figure}

The network was first trained for 200 epochs to an inference accuracy of 87.51$\%$ (Fig. \ref{fig5}a) using SOTA hardware (an NVIDIA RTX 3080 GPU, as before). When simulating deployment of the network weights on our memristive hardware as conducted for MNIST handwritten number classification, we observe $<$0.9$\%$ decrease in accuracy despite all the memristor non-idealities we have implemented in our simulation, with only 20 hardware-aware retraining (HWAT) epochs of the deployed network weights.\\

We implemented device-to-device (DTOD) variation as in our MNIST simulations (Fig. \ref{fig3}). However, to inform future fabrication and estimate the impact of DTOD as well as other device non-idealities on inference accuracy, we varied the DTOD variation values and re-ran simulations for inference in CIFAR-10 (Fig. \ref{fig5}b). The loss in accuracy of $\sim$0.9$\%$ which we report from our CIFAR-10 simulations, is achieved at 30$\%$ DTOD. Due to the low cycle-to-cycle variation, high linearity and good symmetry in our potentiation/depression data, DTOD variation is the dominating factor causing loss in inference accuracy when simulating deployment on our HfO$_x$S$_y$/ HfS$_2$ devices compared to SOTA digital hardware (Fig. \ref{fig5}b). It is important to note that acceptable limits of accuracy loss compared to SOTA hardware are application-dependent. Despite not using a scalable fabrication method in our work, existing literature shows that our image classification accuracy scores are achievable within a realistic DTOD range of 30$\%$ for HfO$_2$-based memristors fabricated from a scalable method (ALD) \cite{HfO2DTODVar}, and for 2D materials-based memristors as well \cite{hBNUltraFastRESETCurveBump}. Similar to our MNIST simulations (Fig. \ref{fig3}c-d), we compare the network's resiliency to drift given a baseline network (uncompensated for drift and without HWAT) and a drift-compensated, HWAT re-trained network (Fig. \ref{fig5}c-d). We observe a much larger variation in accuracy between the HWAT and baseline scores, of $\sim$9$\%$, highlighting the advantage of using HWAT in more challenging machine learning tasks to provide robust weight deployment on analog hardware. With global drift compensation applied, our network modelled on HfO$_x$S$_y$/ HfS$_2$ memristor hardware retains its accuracy of 86.80$\%$ compared to the baseline model without drift compensation and HWAT which degrades in performance significantly over time.\\

\section{Comparison to other memristors}
Evaluating the performance of memristive devices in machine learning applications by simulating the impact of measured device characteristics is a common and important practice in the field. However, different authors use a variety of different tool-kits, such as AIHWKIT \cite{IBMaihwkitRef} \cite{KimEtal}, XPESIM \cite{XPESIM} \cite{FullyHardwareMemrisML}, NeuroSIM \cite{NeuroSim} \cite{NeurSimV2} \cite{lu2021exploring}, and others \cite{lammie2022memtorch} \cite{panetal}. In their simulations, authors deploy various neural network architectures to solve a variety of different machine learning tasks on different datasets, and evaluate different types of memristive device. Therefore, to meaningfully contextualise the performance of our devices, we compare our results to existing literature where potentiation and depression measurements of a few RRAM devices have been used to extract device parameters relevant for weight storage in a neural network. We believe the papers chosen are the most relevant, as Nguyen et al., \cite{nguyen2021memristor} and Lu et al., \cite{lu2021exploring} both employ pulsing schemes with increasing pulse heights in their potentiation and depression experiments. Furthermore, as in our work, Lu et al., \cite{lu2021exploring} also use a chalcogenide switching layer in their Ag/SnS/Pt devices, and do not require electroforming to operate their devices. Pan et al., \cite{panetal} measure devices with a similar structure to ours (TiN/HfO$_2$/Ti), and Yao et al., also use a hafnia-based device stack (TiN/TaO$_x$/HfO$_x$/TiN). It is important to note that the MNIST inference result achieved by Yao et al., was performed fully in hardware, consisting of large memristor arrays connected to integrated programming and read-out circuitry, highlighting a significant achievement in the field \cite{FullyHardwareMemrisML}. However, for inference on the CIFAR-10 dataset, the authors used a neural network with a much larger number of weights (which would require more memristors than they had fabricated). Therefore, for inference on the CIFAR-10 dataset, Yao et al., used a device model which considered the device-to-device and cycle-to-cycle variation which they measured in their experimental hardware \cite{FullyHardwareMemrisML}.\\

\begin{figure}[h!]
\begin{center}
\includegraphics[scale=0.47,trim=6 6 6 6,clip]{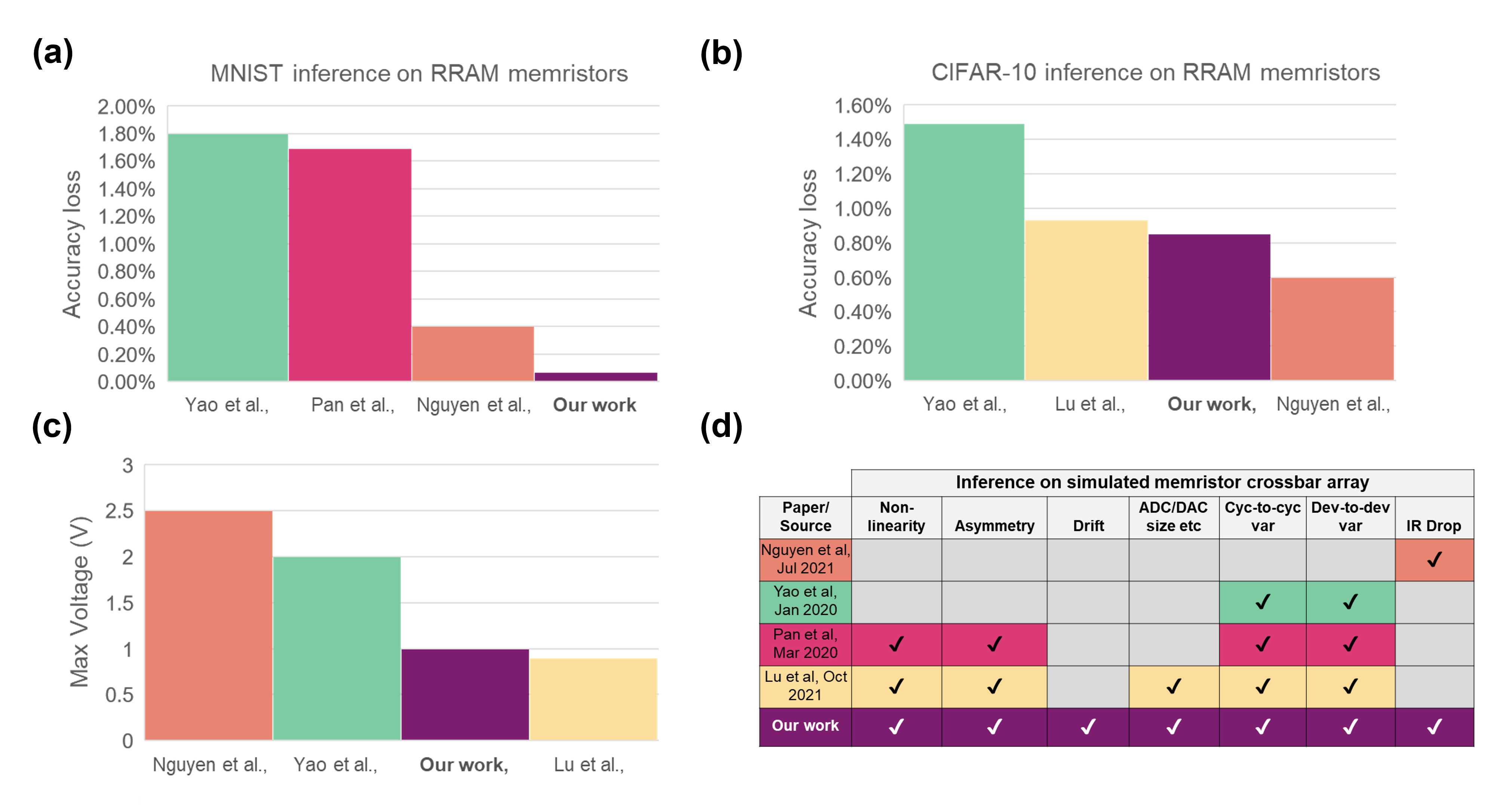}
\end{center}
\caption{Loss in accuracy resulting from deployment of neural networks on simulated analog memristive hardware compared to SOTA, in inference tasks based on test data from the (a) MNIST and (b) CIFAR-10 datasets. Using a simulated crossbar array of our devices, we observe low loss in accuracy for both datasets. (c) Our devices are able to retain high accuracy compared to similar existing literature, despite utilising low programming voltages. (d) To increase the realism of our simulations, we account for a range of device non-idealities while being conscious of peripheral circuitry and drift.} 
\label{fig6}
\end{figure}

In the examples chosen, inference accuracy of a network trained on the MNIST (Fig. \ref{fig6}a) and CIFAR-10 datasets (Fig. \ref{fig6}b) has been evaluated on both SOTA digital hardware and simulated RRAM hardware, allowing for comparison between the two \cite{FullyHardwareMemrisML} \cite{nguyen2021memristor} \cite{lu2021exploring} \cite{panetal}. Absolute accuracy was not used as this depends strongly on the network architecture and size which is not being evaluated here. We also compare the programming voltages used to update memristor weights in CIFAR-10 image classification (Fig. \ref{fig6}c). While there are many other metrics by which the effectiveness of memristive hardware for neural networks can be evaluated, accuracy degradation compared to SOTA in MNIST and CIFAR-10 classification and maximum voltage used during programming were three criteria which were available across a number of different works. Across these metrics, our devices show consistently low loss in accuracy compared to SOTA in both image classification tasks (Fig. \ref{fig6}a-b) while only requiring low programming voltages (Fig. \ref{fig6}c).\\

The table in Fig. \ref{fig6}d indicates the different device non-idealities that have been implemented in each image recognition simulation, disclosed in the main texts or their associated supplementary information. Given all the non-idealities we have implemented in our simulation, we believe our simulations offer a realistic prediction of the performance of arrays of HfO$_x$S$_y$/ HfS$_2$ devices in machine learning tasks. Our devices show promising machine learning performance and strong potential for maintaining high accuracy compared to SOTA hardware, while being programmed with short pulses ($<$350ns) at low voltages ($<$1.0V). Despite the increased computational latency and chip area caused by requiring tailored pulsing schemes, which are used for many memristive devices \cite{NonIdealON/OFFMemris}, including the work by Lu et al., \cite{lu2021exploring}, Nguyen et al., \cite{nguyen2021memristor} and our own, our memristors are forming-free and compliance-free, which contribute towards enabling simplified operation and reduced area consumption for memristor-based chips \cite{memristorareahuang}. 

\section{Conclusions}
We have shown that low energy, fast-switching HfO$_x$S$_y$/ HfS$_2$ memristor hardware can achieve highly linear and symmetric conductance update with high granularity, without requiring electroforming or current compliance. By using the IBM toolkit, we performed highly realistic simulations where not only the real device characteristics are considered but also the impact of a number of other factors such as device-to-device variation, ADCs/DAC size, IR drop and inference accuracy drift over time. The results show that high accuracy is achieved for inference on both the MNIST and CIFAR-10 datasets, showing the potential of resistive memories based on HfO$_x$S$_y$/ HfS$_2$ semiconductor-insulator structures  for future hardware accelerators. With further fine-tuning of device characteristics, (such as operating currents) our forming-free, compliance-free memristors based on HfO$_x$S$_y$/HfS$_2$ have the potential to enable energy-efficient, area-efficient and highly accurate memristor chips for machine learning and neuromorphic computing. \\ 

\section*{Competing interests} 
The authors declare that they have no competing interests.\\

\section*{Data availability}
The data that supports the findings of this study are available from the corresponding author upon reasonable request.\\

\section*{Acknowledgment}
We acknowledge helpful discussions with Abin Varghese, Prabodh Katti and Bipin Rajendran. We acknowledge funding from EP/T517793/1.\\

\bibliography{references.bib}

\begin{thebibliography}{10}

\bibitem{mehonic2022masterplan}
Adnan Mehonic and Anthony~J Kenyon.
\newblock Brain-inspired computing needs a master plan.
\newblock {\em Nature}, 604(7905):255--260, 2022.

\bibitem{jones2018stop}
Nicola Jones et~al.
\newblock How to stop data centres from gobbling up the world’s electricity.
\newblock {\em Nature}, 561(7722):163--166, 2018.

\bibitem{Phys4NeuroComp}
Danijela Markovi{\'c}, Alice Mizrahi, Damien Querlioz, and Julie Grollier.
\newblock Physics for neuromorphic computing.
\newblock {\em Nature Reviews Physics}, 2(9):499--510, 2020.

\bibitem{adnanmemristorsreview}
Adnan Mehonic, Abu Sebastian, Bipin Rajendran, Osvaldo Simeone, Eleni Vasilaki, and Anthony~J Kenyon.
\newblock Memristors—from in-memory computing, deep learning acceleration, and spiking neural networks to the future of neuromorphic and bio-inspired computing.
\newblock {\em Advanced Intelligent Systems}, 2(11):2000085, 2020.

\bibitem{FullyHardwareMemrisML}
Peng Yao, Huaqiang Wu, Bin Gao, Jianshi Tang, Qingtian Zhang, Wenqiang Zhang, J~Joshua Yang, and He~Qian.
\newblock Fully hardware-implemented memristor convolutional neural network.
\newblock {\em Nature}, 577(7792):641--646, 2020.

\bibitem{MNISTDataset}
Yann LeCun, Corinna Cortes, Chris Burges, et~al.
\newblock Mnist handwritten digit database, 2010.

\bibitem{zhu2023hybridhBNCMOS}
Kaichen Zhu, Sebastian Pazos, Fernando Aguirre, Yaqing Shen, Yue Yuan, Wenwen Zheng, Osamah Alharbi, Marco~A Villena, Bin Fang, Xinyi Li, et~al.
\newblock Hybrid 2d--cmos microchips for memristive applications.
\newblock {\em Nature}, 618(7963):57--62, 2023.

\bibitem{NonIdealON/OFFMemris}
Pai-Yu Chen, Binbin Lin, I-Ting Wang, Tuo-Hung Hou, Jieping Ye, Sarma Vrudhula, Jae-sun Seo, Yu~Cao, and Shimeng Yu.
\newblock Mitigating effects of non-ideal synaptic device characteristics for on-chip learning.
\newblock In {\em 2015 IEEE/ACM International Conference on Computer-Aided Design (ICCAD)}, pages 194--199. IEEE, 2015.

\bibitem{DealingMemrisNonIdeal}
Anteneh Gebregiorgis, Abhairaj Singh, Sumit Diware, Rajendra Bishnoi, and Said Hamdioui.
\newblock Dealing with non-idealities in memristor based computation-in-memory designs.
\newblock In {\em 2022 IFIP/IEEE 30th International Conference on Very Large Scale Integration (VLSI-SoC)}, pages 1--6. IEEE, 2022.

\bibitem{lu2021exploring}
Xiu~Fang Lu, Yishu Zhang, Naizhou Wang, Sheng Luo, Kunling Peng, Lin Wang, Hao Chen, Weibo Gao, Xian~Hui Chen, Yang Bao, et~al.
\newblock Exploring low power and ultrafast memristor on p-type van der waals sns.
\newblock {\em Nano letters}, 21(20):8800--8807, 2021.

\bibitem{nguyen2021memristor}
Tien~Van Nguyen, Jiyong An, and Kyeong-Sik Min.
\newblock Memristor-cmos hybrid neuron circuit with nonideal-effect correction related to parasitic resistance for binary-memristor-crossbar neural networks.
\newblock {\em Micromachines}, 12(7):791, 2021.

\bibitem{ZnOformingfree}
Yihui Sun, Xiaoqin Yan, Xin Zheng, Yichong Liu, Yanguang Zhao, Yanwei Shen, Qingliang Liao, and Yue Zhang.
\newblock High on--off ratio improvement of zno-based forming-free memristor by surface hydrogen annealing.
\newblock {\em ACS applied materials \& interfaces}, 7(13):7382--7388, 2015.

\bibitem{SnOFormingCompFree}
Sien Ng, Rohit~Abraham John, Jing-ting Yang, and Nripan Mathews.
\newblock Forming-less compliance-free multistate memristors as synaptic connections for brain-inspired computing.
\newblock {\em ACS Applied Electronic Materials}, 2(3):817--826, 2020.

\bibitem{ReRAMFormingFreeThin}
Binbin Yang, Nuo Xu, Cheng Li, Chenglong Huang, Desheng Ma, Jiahao Liu, Daniel Arum{\'\i}, and Liang Fang.
\newblock A forming-free reram cell with low operating voltage.
\newblock {\em IEICE Electronics Express}, 17(22):20200343--20200343, 2020.

\bibitem{AzizGaSMemristors}
AbdulAziz AlMutairi, Aferdita Xhameni, Xuyun Guo, Irina Chirc{\u{a}}, Valeria Nicolosi, Stephan Hofmann, and Antonio Lombardo.
\newblock Controlled fabrication of native ultra-thin amorphous gallium oxide from 2d gallium sulfide for emerging electronic applications.
\newblock {\em Advanced Materials Interfaces}, page 2400481, 2024.

\bibitem{xhameni2025forming}
Aferdita Xhameni, AbdulAziz AlMutairi, Xuyun Guo, Irina Chirc{\u{a}}, Tianyi Wen, Stephan Hofmann, Valeria Nicolosi, and Antonio Lombardo.
\newblock Forming and compliance-free operation of low-energy, fast-switching hfo$_x$s$_y$/hfs$_2$ memristors.
\newblock {\em Nanoscale Horizons}, 2025.

\bibitem{IBMaihwkitRef}
Malte~J Rasch, Diego Moreda, Tayfun Gokmen, Manuel Le~Gallo, Fabio Carta, Cindy Goldberg, Kaoutar El~Maghraoui, Abu Sebastian, and Vijay Narayanan.
\newblock A flexible and fast pytorch toolkit for simulating training and inference on analog crossbar arrays.
\newblock In {\em 2021 IEEE 3rd international conference on artificial intelligence circuits and systems (AICAS)}, pages 1--4. IEEE, 2021.

\bibitem{UsingIBMaihwkit}
Manuel Le~Gallo, Corey Lammie, Julian B{\"u}chel, Fabio Carta, Omobayode Fagbohungbe, Charles Mackin, Hsinyu Tsai, Vijay Narayanan, Abu Sebastian, Kaoutar El~Maghraoui, et~al.
\newblock Using the ibm analog in-memory hardware acceleration kit for neural network training and inference.
\newblock {\em APL Machine Learning}, 1(4), 2023.

\bibitem{CIFAR10Dataset}
Alex Krizhevsky, Geoffrey Hinton, et~al.
\newblock Learning multiple layers of features from tiny images.
\newblock 2009.

\bibitem{KimEtal}
Yunsur Kim, Hyejin Kim, Seonuk Jeon, Hyun~Wook Kim, Eunryeong Hong, Nayeon Kim, Hyeonsik Choi, Hyoungjin Park, Jiae Jeong, Daeseok Lee, and Jiyong Woo.
\newblock Linear synaptic weight update in selector-less hfo$_2$ rram using al$_2$o$_3$ built-in resistor for neuromorphic computing systems.
\newblock {\em IEEE Transactions on Electron Devices}, 71(8):4637--4643, 2024.

\bibitem{panetal}
Wen-Qian Pan, Jia Chen, Rui Kuang, Yi~Li, Yu-Hui He, Gui-Rong Feng, Nian Duan, Ting-Chang Chang, and Xiang-Shui Miao.
\newblock Strategies to improve the accuracy of memristor-based convolutional neural networks.
\newblock {\em IEEE Transactions on Electron Devices}, 67(3):895--901, 2020.

\bibitem{5bitsForAnalogAI}
Boris Murmann.
\newblock Mixed-signal computing for deep neural network inference.
\newblock {\em IEEE Transactions on Very Large Scale Integration (VLSI) Systems}, 29(1):3--13, 2020.

\bibitem{MlpMnistStructure}
Nipul Manral.
\newblock Mlp-training-for-mnist-classification.
\newblock \url{https://github.com/nipunmanral/MLP-Training-For-MNIST-Classification?tab=readme-ov-file#readme/}, 2019.

\bibitem{HfO2DTODVar}
Eduardo P{\'e}rez, D~Maldonado, C~Acal, JE~Ruiz-Castro, FJ~Alonso, AM~Aguilera, F~Jim{\'e}nez-Molinos, Ch~Wenger, and JB~Rold{\'a}n.
\newblock Analysis of the statistics of device-to-device and cycle-to-cycle variability in tin/ti/al: Hfo2/tin rrams.
\newblock {\em Microelectronic Engineering}, 214:104--109, 2019.

\bibitem{hBNUltraFastRESETCurveBump}
SS~Teja~Nibhanupudi, Anupam Roy, Dmitry Veksler, Matthew Coupin, Kevin~C Matthews, Matthew Disiena, Ansh, Jatin~V Singh, Ioana~R Gearba-Dolocan, Jamie Warner, et~al.
\newblock Ultra-fast switching memristors based on two-dimensional materials.
\newblock {\em Nature Communications}, 15(1):2334, 2024.

\bibitem{ReRAMWan2022}
Weier Wan, Rajkumar Kubendran, Clemens Schaefer, Sukru~Burc Eryilmaz, Wenqiang Zhang, Dabin Wu, Stephen Deiss, Priyanka Raina, He~Qian, Bin Gao, et~al.
\newblock A compute-in-memory chip based on resistive random-access memory.
\newblock {\em Nature}, 608(7923):504--512, 2022.

\bibitem{ResNET1stpaper}
Kaiming He, Xiangyu Zhang, Shaoqing Ren, and Jian Sun.
\newblock Deep residual learning for image recognition.
\newblock In {\em Proceedings of the IEEE conference on computer vision and pattern recognition}, pages 770--778, 2016.

\bibitem{XPESIM}
Wenqiang Zhang, Xiaochen Peng, Huaqiang Wu, Bin Gao, Hu~He, Youhui Zhang, Shimeng Yu, and He~Qian.
\newblock Design guidelines of rram based neural-processing-unit: A joint device-circuit-algorithm analysis.
\newblock In {\em Proceedings of the 56th Annual Design Automation Conference 2019}, pages 1--6, 2019.

\bibitem{NeuroSim}
Xiaochen Peng, Shanshi Huang, Yandong Luo, Xiaoyu Sun, and Shimeng Yu.
\newblock Dnn+ neurosim: An end-to-end benchmarking framework for compute-in-memory accelerators with versatile device technologies.
\newblock In {\em 2019 IEEE international electron devices meeting (IEDM)}, pages 32--5. IEEE, 2019.

\bibitem{NeurSimV2}
Xiaochen Peng, Shanshi Huang, Hongwu Jiang, Anni Lu, and Shimeng Yu.
\newblock Dnn+ neurosim v2. 0: An end-to-end benchmarking framework for compute-in-memory accelerators for on-chip training.
\newblock {\em IEEE Transactions on Computer-Aided Design of Integrated Circuits and Systems}, 40(11):2306--2319, 2020.

\bibitem{lammie2022memtorch}
Corey Lammie, Wei Xiang, Bernab{\'e} Linares-Barranco, and Mostafa~Rahimi Azghadi.
\newblock Memtorch: An open-source simulation framework for memristive deep learning systems.
\newblock {\em Neurocomputing}, 485:124--133, 2022.

\bibitem{memristorareahuang}
Yi~Huang, Takashi Ando, Abu Sebastian, Meng-Fan Chang, J~Joshua Yang, and Qiangfei Xia.
\newblock Memristor-based hardware accelerators for artificial intelligence.
\newblock {\em Nature Reviews Electrical Engineering}, 1(5):286--299, 2024.

\end{thebibliography}
\bibliographystyle{unsrt}

\end{document}